\documentclass{iau_FM}
\usepackage{graphicx}

\title[FM 13.~~Stellar Activity Cycles] %% give here short title %%
{Properties of stellar activity cycles}

\author[Korhonen]   %% give here short author list %%
{Heidi Korhonen$^1$}

\affiliation{$^1$Finnish Centre for Astronomy with ESO, University of Turku, V{\"a}is{\"a}l{\"a}ntie 20, FI-21500 Piikki{\"o}, Finland \\ email: {\tt heidi.h.korhonen@utu.fi}}

\pubyear{2015}
\jname{Astronomy in Focus, Volume 1} 
\editors{Piero Benvenuti, ed.}
\begin{document}

\maketitle

\begin{abstract}
The current photometric datasets, that span decades, allow for studying long-term cycles on active stars. Complementary Ca H\&K observations give information also on the cycles of normal solar-like stars, which have significantly smaller, and less easily detectable, spots. In the recent years, high precision space-based observations, for example from the Kepler satellite, have allowed also to study the sunspot-like spot sizes in other stars. Here I review what is known about the properties of the cyclic stellar activity in other stars than our Sun.
\keywords{stars: activity, chromospheres, late-type, spots, }
\end{abstract}

\firstsection % if your document starts with a section,

\section{Introduction}

The Sun exhibits well established 11-year spot cycle, and a 22-year magnetic cycle. Unfortunately, activity cycles are much more demanding to detect observationally in other stars. Still, long-term monitoring of numerous stars has given us information on the properties of the activity cycles also in other stars. Due to the time and page constraints this review concentrates on observed properties of stellar photospheric and chromospheric cycles. Coronal cycles have also been studied (recent papers include \cite[Lalitha \& Schmitt 2013]{Lalitha13}; \cite[Sanz-Forcada, Stelzer \& Metcalfe 2013]{Sanz-Forcada13}; \cite[Ayres 2015]{Ayres15}).

\section{Methods}

One of the best known methods to study activity cycles is to investigate the chromospheric emission in the cores of the Ca II H\&K lines. This method is based on the fact that the active regions in the solar chromosphere give rise to emission in the cores of the Ca II H\&K lines. Typically, so-called S-index is calculated by comparing the flux in the Ca II H\&K line cores to the flux at a close-by continuum region.

Photospheric activity cycles can also be studied from broad-band photometry. In many active stars the starspots are so large that they cause brightness variations which can be few tens of percent from the mean light level, thus making them easily observable even from the ground. Even smaller spots are reachable from space-based photometry (detectable cycle lengths are limited by relatively short instrument lifetimes). Small ground-based automatic photospheric telescopes, that have been operational last two decades or longer, provide excellent data-sets for cycle studies.

Doppler imaging (see, e.g., \cite[Vogt \etal\ 1987]{Vogt87}; \cite [Piskunov \etal\ 1990]{Piskunov90}) is a method that provides the best spatial resolution on the stellar surface. It uses high resolution, high signal-to-noise spectroscopic observations obtained at different rotational phases of the star. If the star has a non-uniform surface temperature, i.e., has starspots, the spectral lines show small distortions from the normal Gaussian shape. These distortions move in the line-profile when the position of the starspots on the surface changes, due to the change of line-of-sight velocity caused by the stellar rotation. The temperature maps, which are constructed by tracking the movement of these distortions, can be used for detailed studies of stellar spot configurations. If spectropolarimetric observations are obtained over the stellar rotation, similarly to what is done in Doppler imaging, the surface magnetic field of the star can mapped using Zeeman-Doppler imaging technique (\cite[Semel 1989]{Semel89}).

\section{Activity cycles in 'normal' stars}

One cannot talk about stellar activity cycles without mentioning the Mt. Wilson H\&K survey. The project was initiated in 1966 by Olin Wilson (\cite[Wilson 1968]{Wilson68}; \cite[Wilson 1978]{Wilson78}), and it originally monitored 91 stars with spectral types ranging from F to M. The monitoring was expanded to more stars and star types, and in total some 2000 stars were observed during the project's lifetime (\cite[Baliunas \etal\ 1995]{Baliunas95}; \cite[Baliunas \etal\ 1998]{Baliunas98}). 

The project found many stars with variations in the chromospheric emission, and distinguished four different activity types: variable, cyclic, trend, and flat activity. From the core sample of 111 dwarf stars 60\% of the targets show cyclic or apparently cyclic behaviour, 25\% are variable and 15\% exhibit flat activity (\cite[Baliunas \etal\ 1995]{Baliunas95}). In the sample of 175 more evolved stars, 40\% show cyclic and basically the same fraction variable activity (\cite[Baliunas \etal\ 1998]{Baliunas98}). Meaning that the regularly cyclic activity seems to change to more chaotic behaviour when the star evolves.

For correlating the photospheric and chromospheric activity one can compare the chromospheric indicators from Ca II H\&K and the photospheric spots from broadband photometry. \cite[Radick \etal\ (1998)]{Radick98} compared these two different indicators. Their results show that the younger, more active stars tend to become fainter as their HK emission increases, whereas the older stars tend to become brighter as their HK emission increases. This means that the activity of young stars is spot dominated, and that of the older stars is dominated by photospheric faculae.

\cite[Brandenburg, Saar, \& Turpin (1998)]{Brandenburg98} correlated the stellar activity with $\log \omega_{cyc}/\Omega$ and noticed that the stars concentrate on ‘active’ and ‘inactive’ branches, later also ‘super saturated’ branch was added (\cite[Saar \& Brandenburg 1999]{Saar99}). Stars move along the active branch as they age. Once reaching activity range -4.8 -- -4.7 the star makes a rapid, maybe discontinuous, transition to the inactive branch (see also \cite[Saar \& Brandenburg 2002]{Saar02}). Several authors have correlated cycle length with the rotation period (e.g., \cite[Baliunas \etal\ 1996]{Baliunas96}; \cite[B{\"o}hm-Vitense 2007]{Böhm07}; \cite[Ol{\'a}h \etal\ 2009]{Olah09}). Stars are seen to group again to the inactive and active branches, and longer rotation periods tend to produce longer cycles

The original Mt. Wilson sample does not include many M dwarfs. \cite[Gomes da Silva \etal\ (2012)]{Gomes12} studied 27 M0 -- M5.5 dwarfs observed by HARPS planet search. Half of the sample showed significant cycles in the time range of few years. Surprisingly, there does not seem to be any correlation between cycle length and rotation period for M dwarfs (\cite[Savanov 2012]{Savanov12}).

\section{Cycles in active stars}

In many active stars the starspots are so large that they cause brightness variations which can be few tens of percent from the mean light level, thus making them easily observable even from the ground. Many studies on photospheric cycles in active stars have been carried out, e.g., \cite[Henry \etal\ (1995)]{Henry95}, \cite[Jetsu (1996)]{Jetsu96}, \cite[Ol{\'a}h, Koll{\'a}th \& Strassmeier (2000)]{Olah00}, \cite[Lanza \etal\ (2002)]{Lanza02}, \cite[Berdyugina, Pelt \& Tuominen (2002)]{Berdyugina02}, \cite[Donati \etal\ (2003)]{Donati03}, \cite[J{\"a}rvinen \etal\ 2005]{Jarvinen05}, \cite[Savanov (2009)]{Savanov09}, \cite[Vida \etal\ (2010)]{Vida10}, \cite[Lehtinen \etal\ (2012)]{Lehtinen12}, and \cite[Metcalfe \etal\ (2013)]{Metcalfe13}. Here few papers are discussed in more detail as examples of typical results. 

\cite[Messina \& Guinan (2002)]{Messina02} monitored six K0 - G0 active dwarfs stars for more than a decade. Many of their main results are similar to the ones obtained from the more extended Mt. Wilson sample of stars with more 'normal' activity levels. \cite[Messina \& Guinan (2002)]{Messina02} find that the observed cycle lengths seem to converge with stellar age from a maximum dispersion around the Pleiades' age towards the solar cycle value at the Sun's age, and that the overall short- and long-term photometric variability increases with inverse Rossby number.

The cycles active stars show are not as regular and cyclic as many of their more normal counterparts have. \cite[Ol{\'a}h, Koll{\'a}th \& Strassmeier (2000)]{Olah00} already show that many active stars exhibit multiple cycle lengths simultaneously. The more detailed analysis using short-term Fourier transforms shows that the cycle lengths in active stars are also often variable (\cite[Ol{\'a}h \etal\ 2009]{Olah09}). The positive relation between the rotational period and cycle length is also seen from the \cite[Ol{\'a}h \etal\ (2009)]{Olah09} sample.

Active stars also show more exotic cycles, e.g., the so-called flip-flop phenomenon. This effect was first discovered in an active single giant, FK Com, in the early 1990’s (\cite[Jetsu \etal\ 1993]{Jetsu93}). In this phenomenon the activity concentrates on two permanent active longitudes, and flips between the two every few years. Since the discovery this effect has been reported in binaries (e.g., \cite[Berdyugina \& Tuominen 1998]{Berdyugina98}), young active stars (e.g., \cite[J{\"a}rvinen \etal\ 2005]{Jarvinen05}), and even in the Sun (e.g., \cite[Berdyugina \& Usoskin 2003]{Berdyugina03}). Recent studies show that flip-flops are not as regular and cyclic as at some point thought, and that there are also smaller ‘phase jumps’ (e.g., \cite[Korhonen, Berdyugina \& Tuominen 2002]{Korhonen02}; \cite[Ol{\'a}h \etal\ 2006]{Olah06}; \cite[Hackman \etal\ 2013]{Hackman13}).

\section{Butterfly diagrams}

It would be very important to obtain information on the spot latitudes during activity cycles in other stars. In the Sun we know that the first spots of the cycle appear at higher latitudes ($\pm 30^{\circ}$) and migrate towards the equator as the cycle advances. We assume the same behaviour in the other stars, but we have not really produced proper 'butterfly diagrams' of distant stars. This would be best done using Doppler imaging, where we get detailed information on the spot latitudes. Still, it is difficult to obtain frequent enough long-term data for these studies.

Period variations have been detected in many active stars (e.g. \cite[Messina \& Guinan 2003]{Messina03}; \cite[Vida \etal\ 2014]{Vida14}). These result can be explained by surface differential rotation and changing spot latitudes. Still, from these photometric studies we do not obtain accurate latitude information, and cannot tell which direction the spots have migrated. \cite[Berdyugina \& Henry (2007)]{Berdyugina07} recovered the butterfly diagram of active RS CVn binary HR~1099 from photometry. They used the known differential rotation law, obtained from previous Doppler imaging work, in their reconstruction.

\section{Magnetic cycles}

Spectropolarimetric observations of stars and mapping the detailed configurations of the surface field has been possible soon for two decades. Still, the weak signal from the magnetic field demands superb observations, and the technique of Zeeman Doppler imaging requires many observations over a rotational phase for accurate surface reconstruction. Therefore, the observations of the stellar magnetic cycles have become possible only recently.

\begin{figure}[b]
% \vspace*{-2.0 cm}
\begin{center}
 \includegraphics[width=2.5in]{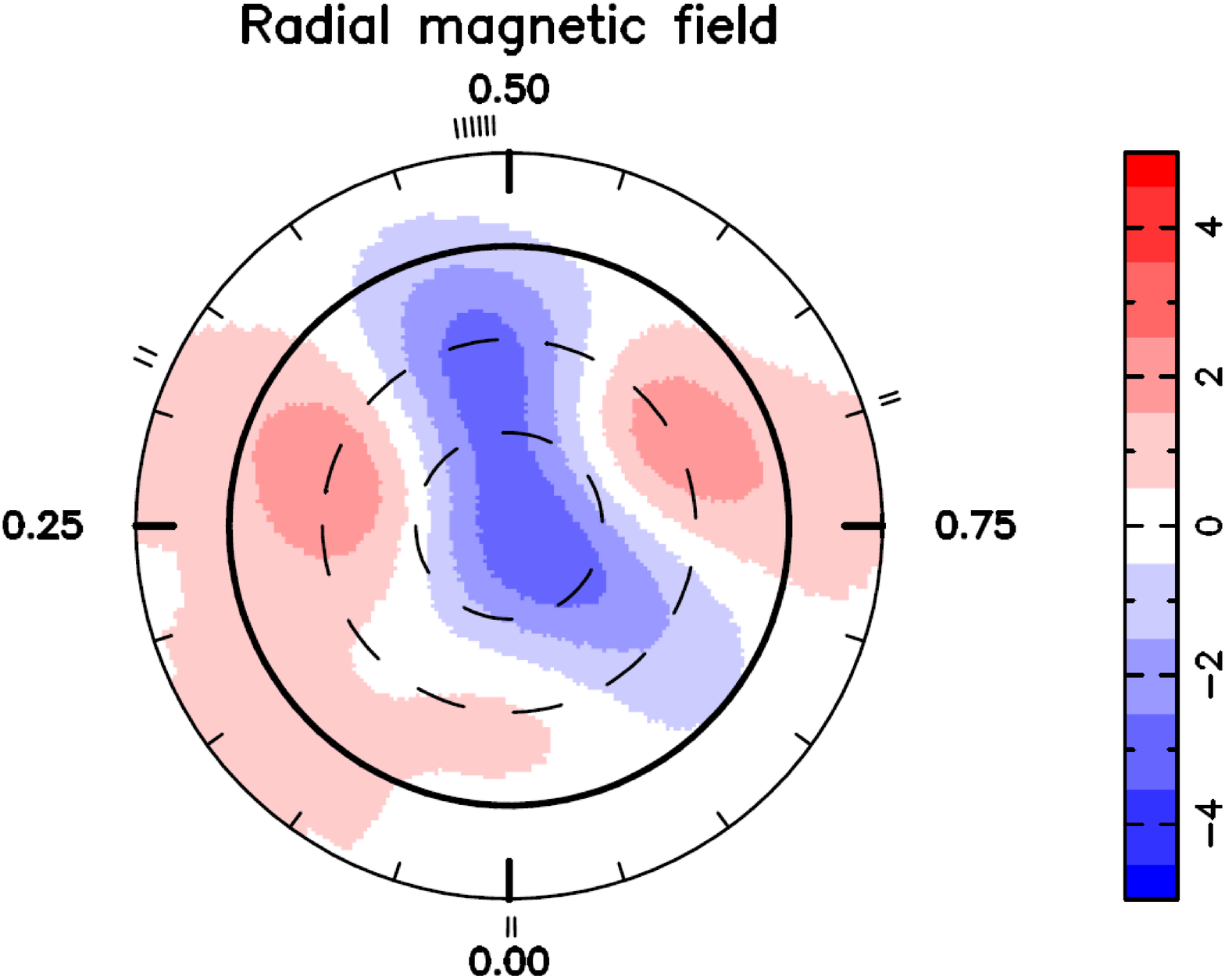}
 \includegraphics[width=2.5in]{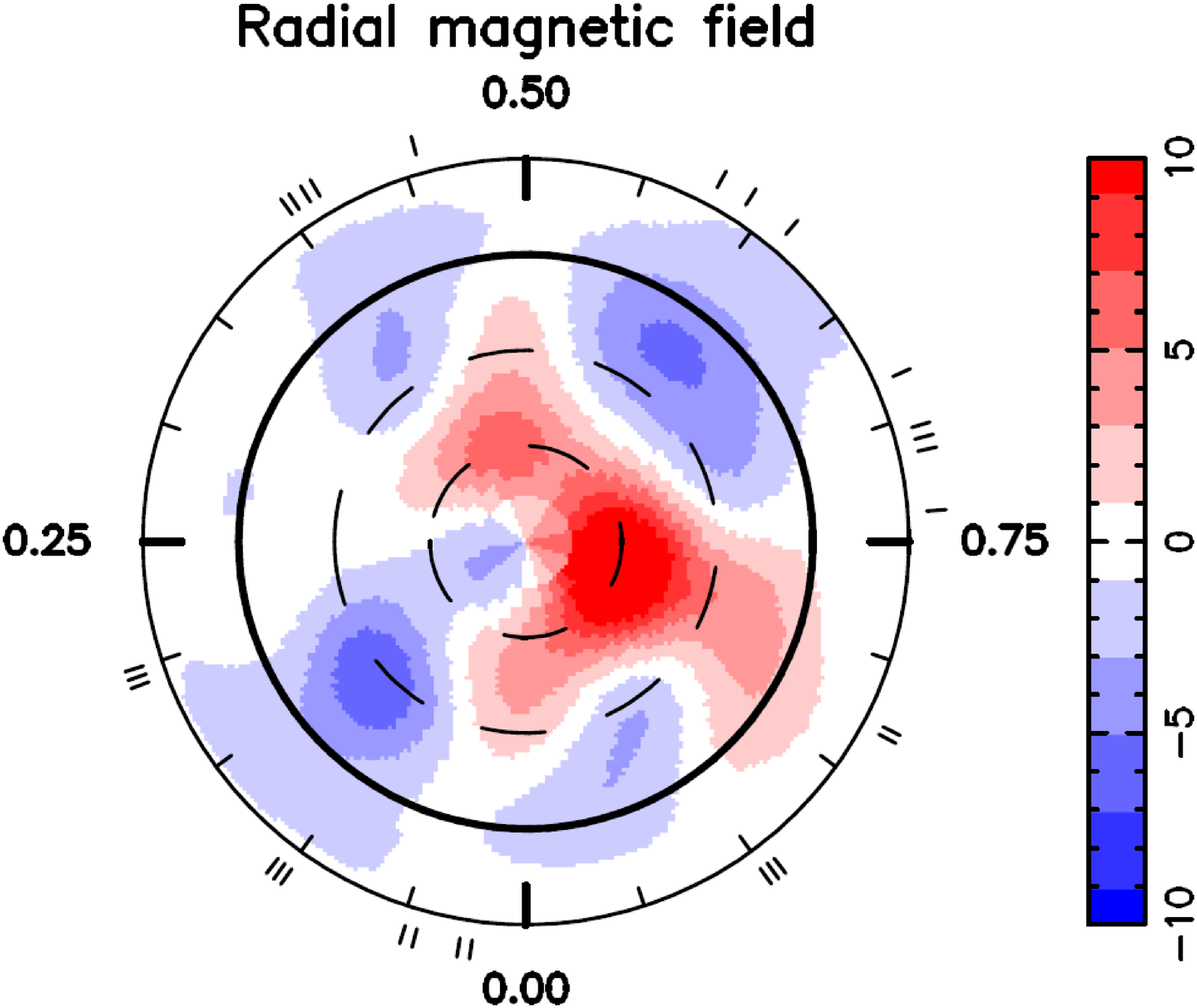}
% \vspace*{-1.0 cm}
 \caption{Radial magnetic field of $\tau$~Boo in June 2006 (left, from \cite[Catala \etal\ 2007]{Catala07}) and in June-July 2007 (right, from \cite[Donati \etal\ 2008]{Donati08}. The maps show the reversal of the polarities in the polar area.}
   \label{fig1}
\end{center}
\end{figure}

The first polarity revels in another star than the Sun was observed in the planet hosting F7 dwarf $\tau$~Boo (\cite[Donati \etal\ 2008]{Donati08}). This was established by comparing the magnetic surface maps from June 2006 (\cite[Catala \etal\ 2007]{Catala07}) and June-July 2007 (\cite[Donati \etal\ 2008]{Donati08}), which are shown in Fig.~\ref{fig1}. In 2008 the second reversal in $\tau$~Boo was seen by Fares et al. (2009), implying a magnetic cycle of about 2 years.

\cite[Petit \etal\ (2009)]{Petit09} discovered a polarity reversal in rapidly rotating solar-like star HD190771. But so far no cycle has been detected. Other long-term monitoring of the magnetic field configurations include e.g., K2 dwarf $\epsilon$~Eri (\cite[Jeffers \etal\ 2014]{Jeffers14}), rapidly rotating solar-like star $\zeta$~Boo (\cite[Morgenthaler \etal\ 2012]{Morgenthaler12}), and a sample of solar-like stars (\cite[Morgenthaler \etal\ 2011]{Morgenthaler11}). No clear periodic reversals have been found in these studies, but we are collecting a dataset of long-term observation which will enable studying magnetic cycles in other stars than the Sun.

\section{Final remarks}

Since the discovery of activity cycles, first in our own Sun and later in other stars, we have identified many characteristics of these cycles. 

Stellar activity cycles are seen in a large number of stars in the lower main sequence, meaning stars that have masses approximately solar mass and less. Several studies have also shown that the stellar rotation period and the cycle length correlate, i.e., stars with longer rotation period produce longer cycles. Therefore also the cycle length changes with stellar age, while the stellar rotation rate decreases due to the magnetic breaking. Very active stars, with huge starspots, can have multiple and changing activity cycles, and also exhibit ‘exotic’ behaviour: active longitudes, phase jumps and flip-flops. 

Longer spectropolarimetric monitoring and Zeeman-Doppler imaging techniques are providing the opportunity to also probe the full magnetic field configuration changes during stellar cycles. So far stellar magnetic cycle has been seen in one stars, tau Boo, but the polarity change in the polar regions has seen in few others. 

With dedicated telescopes and improved observing techniques we are starting to probe the detailed properties of stellar activity cycles. This is an important development as the cyclic magnetic activity is produced by Dynamo action operating in the star, and by studying the properties of the cycles, we can also give constraints to the Dynamo operation itself.

{\bf Acknowledgments}
The author acknowledges the travel support from Emil Aaltonen foundation, Turku University Foundation, and IAU to participate the IAU General Assembly 2015.

\end{document}